\documentclass[twocolumn,prl,aps,nobalancelastpage,amssymb]{revtex4}

\usepackage{graphicx}

\def \Pr124{PrBa$_2$Cu$_4$O$_8$}
\def \beqn {\begin{equation}}
\def \bfig {\begin{figure}}
\def \btab {\begin{table}}
\def \eeqn {\end{equation}}
\def \efig {\end{figure}}
\def \etab {\end{table}}

\begin{document}

\title{Possible co-existence of local itinerancy and global localization in a quasi-one-dimensional conductor}

\author{A. Narduzzo A. Enayati-Rad, S. Horii$^\dagger$ \& N. E. Hussey}

\affiliation{H. H. Wills Physics Laboratory, University of Bristol, Tyndall Avenue, BS8 1TL, U.K.}

\affiliation{$^\dagger$Department of Applied Chemistry, Tokyo University, 7-3-1 Hongo, Tokyo, 113-8656, Japan}

\date{\today}

\begin{abstract}
In the chain compound \Pr124 localization appears simultaneously with a dimensional crossover in the electronic ground
state when the scattering rate in the chains exceeds the hopping rate between the chains. Here we report the discovery
of a large, transverse magnetoresistance (T-MR) in \Pr124 in the localized regime. This result suggests a novel form of
localization whereby electrons retain their metallic (quasi-one-dimensional) character over a microscopic length scale
despite the fact that macroscopically, they exhibit localized (one-dimensional) behavior.
\end{abstract}

\maketitle

The physics of purely one-dimensional (1D) conducting chains is a rich and fascinating area of solid state research due
primarily to the emergence of exotic ground states and the availability of exact solutions \cite{GiamaBook, Voit}. Once
coupling is introduced between chains however, as in real electronic systems, our understanding is far from complete,
especially when disorder and/or interactions are considered. While it is established that all states are localized in
strictly 1D systems containing arbitrarily weak disorder \cite{MottTwose, Ishii, KaneFisher, Abrahams}, in quasi-1D
conductors, the onset of localization \cite{AbrikosovRyshkin, PrigodinFirsov} and the transition to the
Tomonaga-Luttinger liquid (TLL) state \cite{ Castellani, Kopietz} remain theoretical controversies, prompting the
search for chain-containing compounds whose effective dimensionality can be controlled by tuning the interchain
coupling.

\begin{figure}
\includegraphics[width=6.0cm,keepaspectratio=true]{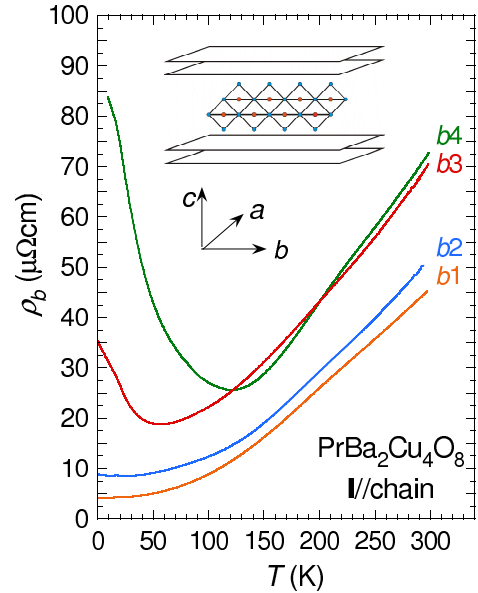}
\caption{Zero-field $\rho_b$($T$) curves of four Pr124 crystals with different levels of disorder. Inset: schematic of
the Pr124 crystal structure.} \label{Figure1}
\end{figure}

\Pr124 (Pr124), with its network of double CuO chains sandwiched between insulating CuO$_2$ bilayers (see Fig.
\ref{Figure1}), is one such system. In clean Pr124 crystals, a highly anisotropic 3D Fermi-liquid state develops at low
temperatures with $\rho$($T$) $\sim T^2$ along all three crystallographic axes ($\rho_a$:$\rho_b$:$\rho_c$ $\sim$
1000:1:3000) \cite{Hussey02, McBrien02}. Due to the small interchain hopping energies (2$t_{a,c} \sim$ 3-5meV),
coherent interchain tunneling is extremely fragile to external perturbations and dimensional crossover phenomena have
been observed both as a function of temperature \cite{McBrien02} and magnetic field $H$ \cite{Hussey02}. It was also
shown recently that when the residual in-chain resistivity $\rho_{b0}$ exceeds a certain threshold value, $\rho_b$($T$)
develops an upturn at low $T$\cite{Narduzzo}. This upturn occurs in a regime where the intrachain elastic scattering
rate $\hbar/\tau_0 > 2t_{a,c}$, suggesting a close correlation between interchain coherence and localization of
in-chain states. The possibility to vary both dimensionality and the nature of the metallic ground state makes Pr124 a
potentially ideal system with which to resolve experimentally the controversies outlined above.

In this Letter, we probe the in-chain charge dynamics in Pr124 across the localization threshold via a detailed and
systematic in-chain magnetotransport study. In the absence of any charge-density-wave (CDW) formation, a unique form of
localization is implied, whereby the chain network becomes fragmented into metallic islands sandwiched between strong
back-scattering impurities. In contrast to existing models \cite{PrigodinFirsov, KuseZeller, RiceBernasconi} however,
the data imply that locally, finite coherent interchain hopping survives despite the fact that globally, $\hbar/\tau_0$
$>$ 2$t_{a,c}$.

All four needle-like crystals used in this study (and in Ref. \cite{Narduzzo}) were grown by a self-flux method in MgO
crucibles under high oxygen pressures \cite{Horii00}. Elemental analysis revealed the presence of Mg as the most
abundant impurity element with a content that scaled approximately with the (extrapolated) residual resistivities. The
$\rho_b$($T$,$H$) measurements were carried out for 1.4K $\leq T < $ 300K using a standard four-probe ac lock-in
technique in a 14 Tesla magnet. For a detailed description of the mounting and measurement procedure, in particular the
isolation of the in-chain current response and the absolute determination of $\rho_b$, please see Ref. \cite{McBrien02,
Narduzzo}. Our estimate of the uncertainty in the stated values is $\pm 20\%$.

Fig. \ref{Figure1} shows $\rho_b$($T$) of the four crystals, labelled {\it b}1-{\it b}4 in order of increasing
resistivity at $T$ = 300K. Only $b1$ remains metallic down to $T$ = 0.3K. All other crystals have a resistivity minimum
at $T$ = $T_{\rm min}$. For sample $b2$, close to the localization threshold ($T_{\rm min}$ = 27K), the nominal
zero-temperature mean-free-path $\ell_0$ $\sim$ 200$\AA$ equivalent to $k_F \ell_0 \sim$ 50 ($k_F$ being the in-chain
Fermi wave vector). The kinks observed in $\rho_b$($T$) in samples $b2-4$ are due to N\'{e}el ordering of the Pr ions
at $T_{\rm N}$(Pr) = 17.5K. Since Pr ordering occurs in {\it all} crystals (with or without upturns) at around the same
temperature, the resistivity upturns, that appear only in more disordered crystals and at varying $T_{\rm min} > T_{\rm
N}$(Pr), are clearly of a different origin.

\begin{figure}
\includegraphics[width=8.5cm,keepaspectratio=true]{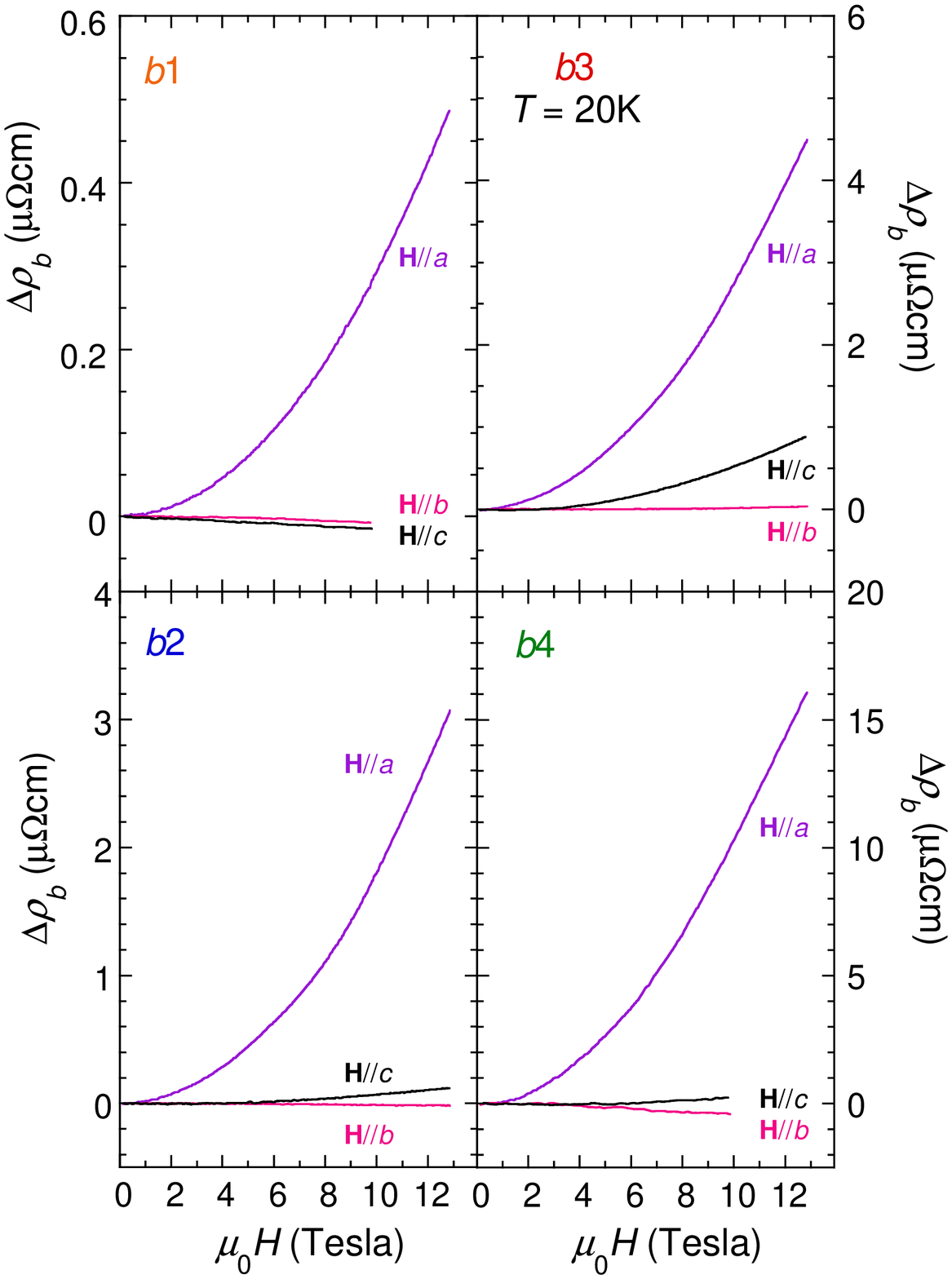}
\caption{Magnetoresistance sweeps at $T$ = 20K for all four crystals. Each panel ($b1$ to $b4$) shows $\Delta \rho_{b}$
for {\bf H} aligned along the three different crystallographic axes.} \label{Figure2}
\end{figure}

The panels of Fig. \ref{Figure2} display the change in resistivity $\Delta \rho_b = \rho_{b}(\mu_0H)-\rho_b(0)$ as a
function of applied field for $b1$-$b4$ at $T$ = 20K \cite{Narduzzo2}. Let us consider first the metallic sample $b1$:
the T-MR for {\bf H}//$a$ is large, positive and varies quadratically with $\mu_0H$ up to 13T. By contrast, the T-MR
for {\bf H}//$c$ is small, negative and similar in size to the longitudinal MR. This striking anisotropy in the MR
response indicates that it is related to carriers on the CuO chains rather than the insulating planes. The marked
difference in the T-MR for {\bf H}//$c$ and {\bf H}//$a$ is surprising though given the Fermi surface topology and the
fact that the warping of the quasi-1D Fermi sheets is comparable in both interchain directions \cite{Hussey02}. One
possible explanation for this anisotropy is that the orientation of the zig-zag double chain network along the $c$-axis
(see Fig. \ref{Figure1}) gives rise to a larger hopping integral within the bi-chains and hence a larger orbital MR for
{\bf H}//$a$. This conjecture is not consistent however with the observation of a {\it large} T-MR for {\bf I}//$a$ and
{\bf H}//$c$ \cite{Horii02}. The more likely origin for the T-MR anisotropy for {\bf I}//$b$ is a near-perfect
cancellation of the magnetoconductance $\Delta\sigma$ for {\bf H}//$c$ by the Hall conductivity $\sigma_{xy}$. The
general expression for {\bf H}//$z$ is $\Delta \rho_{xx}/\rho_{xx}(0)$ = - $\Delta \sigma_{xx}/\sigma_{xx}(0)$ -
$\sigma_{xy}^2/\sigma_{xx}(0)\sigma_{yy}(0)$. In our case, $x=b$ and $y=a(c)$ for {\bf H}//$c(a$). In a perfectly
isotropic metal, the two terms in the sum are equivalent and correspondingly, the T-MR is zero. In an anisotropic
(quasi-1D or quasi-2D) metal, such cancellation is also possible, but only for in-chain(plane) currents and isotropic
mean-free-paths. The observation of finite T-MR for {\bf H}//$a$ requires the condition $\sigma_{ba} > \sigma_{bc}$ and
thus some additional momentum-dependence in the integral for $\sigma_{ba}$ that is absent for $\sigma_{bc}$.

\begin{figure}
\includegraphics[width=7.0cm,keepaspectratio=true]{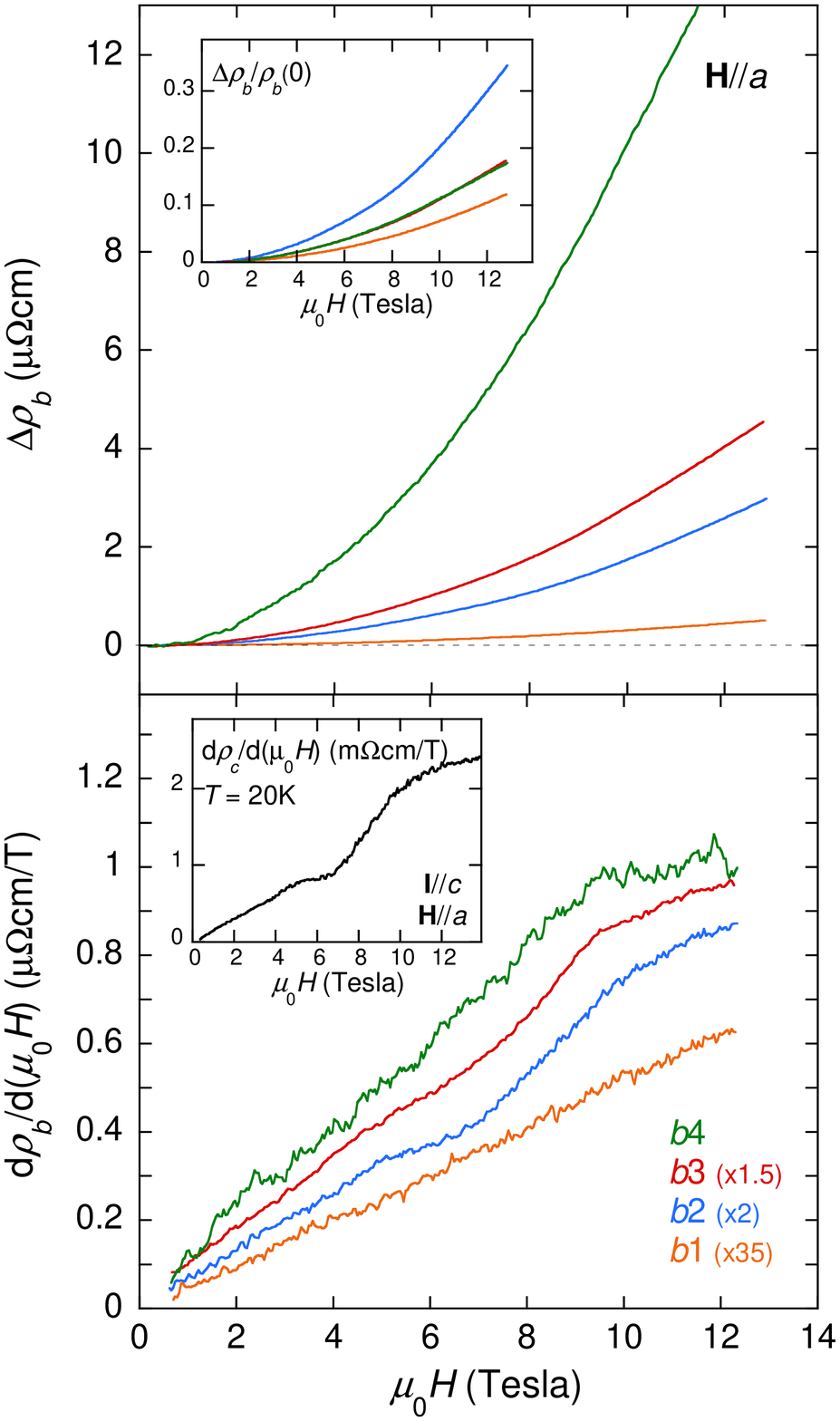}
\caption{Comparison of MR sweeps at $T = 20$K for {\bf H}//$a$. Top panel: $\Delta \rho_{b}$ vs. $\mu_0H$. Inset to top
panel: $\Delta \rho_b$/$\rho_b$(0) vs. $\mu_0H$. Bottom panel: Corresponding $d\rho_{b}/d(\mu_0H)$ vs. $\mu_0H$, scaled
for clarity. Inset to bottom panel: $d\rho_{c}/d(\mu_0H$) vs. $\mu_0H$ for a {\it metallic} crystal with $\ell_0 \sim
600\AA$ at $T$ = 20K (from Ref.~\cite{Hussey02}).} \label{Figure3}
\end{figure}

The most surprising feature of Fig. \ref{Figure2} is the observation of similarly strong anisotropy in the MR response
of the more disordered samples. In the top panel of Fig. \ref{Figure3} we compare directly $\Delta \rho_b$($\mu_0H$) at
$T$ = 20K and {\bf H}//$a$ for each crystal. Remarkably, $\Delta \rho_b$($\mu_0H$) for $b$4 is more than one order of
magnitude larger than for $b$1. Indeed, the T-MR is smallest in $b1$ {\it even when $\Delta \rho_b$($\mu_0H$) is scaled
by the corresponding zero-field value $\rho_b$(0)}, as shown in the inset to the top panel of Fig. \ref{Figure3}.

Before discussing possible origins of the large T-MR in the more disordered crystals, let us first highlight two
aspects of their T-MR response that differ from that of the metallic sample; namely the field dependence and Kohler's
scaling. The bottom panel of Fig. \ref{Figure3} shows d$\rho_b$/d($\mu_0H$) for $b$1-4 at $T$=20K (scaled for clarity).
Below 5T, the derivative is linear (i.e. $\Delta \rho_b \propto H^2$) for all four crystals and remains so for $b$1 at
all field strengths studied. In the three other samples however, d$\rho_b$/d($\mu_0H$) shows kinks around 5T and 10T.
These kinks are only observed for {\bf H}//$a$ and are present above and below both $T_{\rm min}$ and $T_{\rm N}$(Pr).
This peculiar field-dependence in fact mirrors that observed for {\bf I}//$c$ and {\bf H}//$a$ in clean crystals that
remain metallic at low $T$ \cite{Hussey02}. We reproduce one such curve (at $T$ = 20K) in the inset to the bottom panel
of Fig. \ref{Figure3} for comparison. For {\bf I}//$c$, these two kinks are attributed to a field-induced
renormalization of the $c$-axis warping and provide estimates for 2$t_c$ on the individual Fermi sheets
\cite{Hussey02}. The increased sensitivity of the (conduction) electrons to small perturbations is probably an effect
due to weaker screening of the carriers in a state of lower dimensionality (i.e. once $\hbar/\tau_0 > 2t_{a,c}$) and is
also manifest in the appearance of the kink in $\rho_b$($T$) of the more disordered crystals at $T_{\rm N}$(Pr) (see
Fig. \ref{Figure1}), a feature that is universally observed in $\rho_c$($T$) \cite{McBrien02}.

\begin{figure}
\includegraphics[width=7.0cm,keepaspectratio=true]{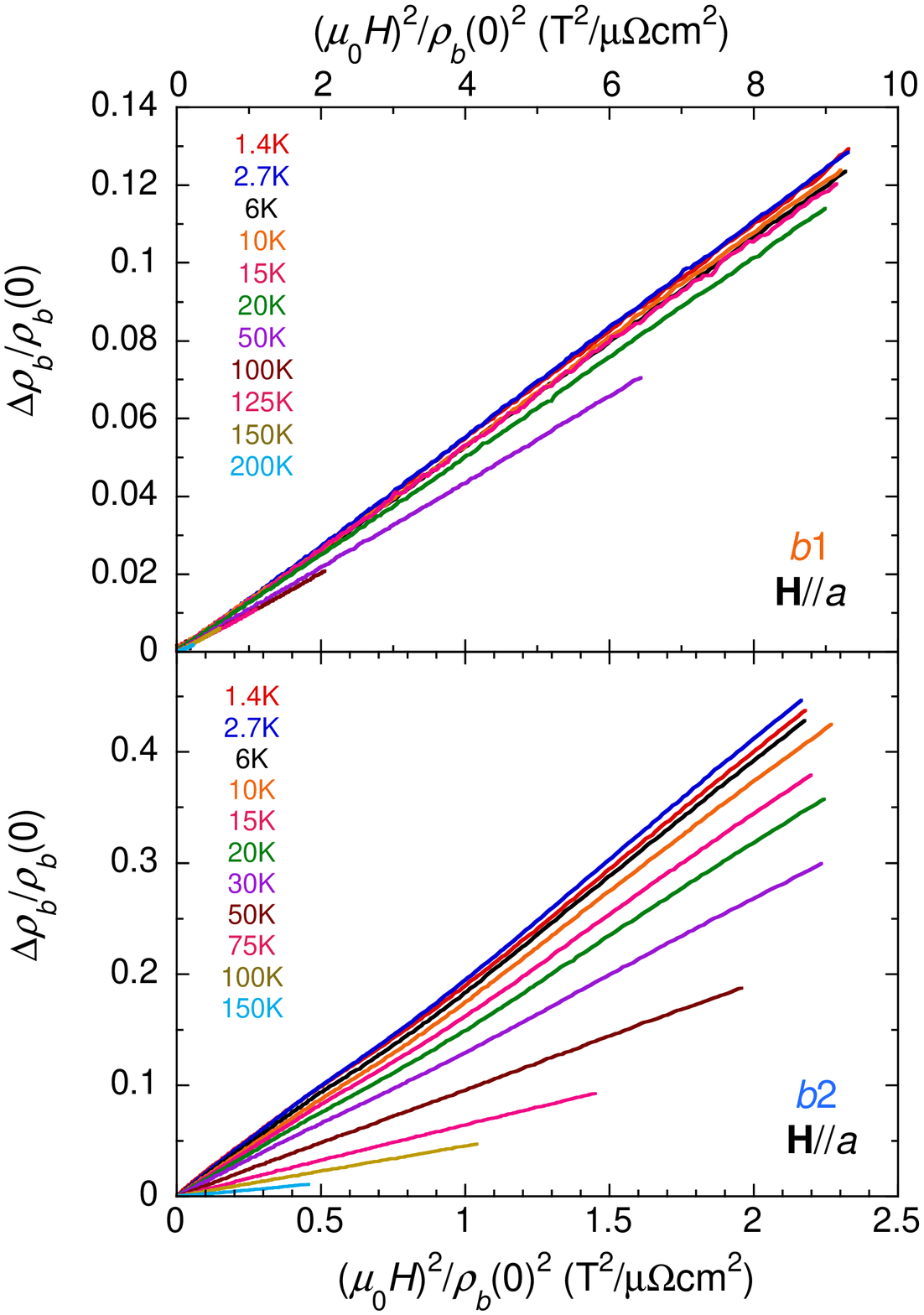}
\caption{Kohler's scaling for sample $b1$, above, and $b2$, below at various temperatures 1.4K $\leq T \leq$ 200K.}
\label{Figure4}
\end{figure}

Kohler's rule states that $\Delta \rho/\rho\propto(H/\rho)^2$ for $\Delta \rho(H)\propto H^2$. Thus by plotting $\Delta
\rho$/$\rho$(0) vs. ($H/\rho$(0))$^2$, all data should collapse onto a single curve independent of temperature. Such
scaling, which is only observed in metallic systems, implies that the lifetime controlling the cyclotron motion is the
same as that which determines $\rho(T)$ \cite{AbrikosovBook}. Fig. \ref{Figure4} shows Kohler plots for samples $b1$
and $b2$. For $b1$, Kohler's scaling is reasonably well obeyed below 20K with small deviations from full scaling
thereafter. In contrast, for $b2$, Kohler's rule is violated at all $T$ (in fact the product $\Delta \rho_b.\rho_b(0)$
decreases exponentially with $T$). This difference in Kohler's scaling between $b1$ and $b2$ is striking given the very
similar $\rho_b$($T$) behavior of the two crystals (Fig. \ref{Figure1}) and suggests that a Boltzmann-type approach
with its associated relaxation time approximation is no longer valid once $\hbar/\tau_0$ $\geq$ $2t_{a,c}$. Such
violation of Kohler's rule is reminiscent of that seen in the high-$T_c$ cuprates where it has been attributed to the
presence of two scattering times possibly associated with spin-charge separation, a hallmark of the TLL state
\cite{Harris95}.

The presence of such a large positive T-MR appears to rule out Kondo scattering as the origin of the low-$T$ upturn in
$\rho_b$($T$) in Pr124. A positive quadratic MR may occur in localized quasi-1D compounds due to strong spin-orbit
scattering (anti-localization) \cite{DougdaleBook} or field-enhanced spin-flip scattering \cite{MartinPhillips} but
these effects tend to be isotropic and in the case of the latter, scale as ($H$/$T$)$^2$ \cite{MartinPhillips}, in
marked contrast to what is observed here. Finally, neither the $T$-dependence of $\rho_b$ nor the $H$-dependence of
$\Delta \rho_b$ are compatible with variable-range-hopping \cite{VRH}.

In partially-gapped CDW compounds such as NbSe$_3$ \cite{Coleman} and $A$Mo$_6$O$_{17}$ ($A$ = K, Tl, Na) \cite{Tian},
a large positive MR is observed for fields perpendicular to the principal conduction axis due to field-enhanced nesting
of the Fermi sheets \cite{BalseiroFalicov} and resultant two-band effects which lead to a violation of Kohler's rule
\cite{Tian, Shen03}. One might consider a similar process occurring in Pr124 whereby disorder progressively pins the
slowly fluctuating charge order \cite{Fujiyama03}. It is well known however that disorder also tends to randomize the
phases of the CDW and so disturb the coherence needed for achievement of the phase transition. Moreover this scenario
is difficult to reconcile with the fact that $\Delta \rho_b$($H$) in samples $b$2-4 is qualitatively the same as that
seen for {\bf I}//$c$ and {\bf H}//$a$ in crystals showing no resistivity upturn and therefore no pinned CDW
\cite{Hussey02}. In the latter, the low-field T-MR is entirely consistent with Boltzmann transport theory. We conclude
therefore that whilst the electronic states in disordered Pr124 are showing clear signatures of localization, $\Delta
\rho_b$($H$) has a similar {\it orbital} origin to that seen in the low-$T$ metallic state; the enhanced T-MR and
violation of Kohler's scaling in the more disordered crystals resulting from a change in dimensionality and subsequent
re-balancing of the two T-MR components $\Delta\sigma$ and $\sigma_{xy}$ \cite{Harris95}.

The essential point here is that even though samples $b$2-4 show localized behavior, their nominal $\ell_0$ values
(i.e. at low $T$) are still very large ($>$ 200$\AA$ for $b2$ and $\sim 100 \AA$ for $b3$ and $b4$, estimated by
extrapolating $\rho_b$($T$) from the metallic side to 0K). In Pr124, localization develops once the impurity-dominated
scattering rate $\hbar/\tau_0 \sim 2t_{a,c}$, due to the loss of long-range {\it interchain} coherence, rather than at
$\hbar/\tau_0 \sim 2t_b$, the usual Mott-Ioffe-Regel criterion. This implies that the (disorder-induced) dimensional
crossover has a profound effect on the impurity states themselves and induces singular (strong back-scattering) effects
akin to impurity states in 1D \cite{KaneFisher}.

The main finding of our MR study though is that once $\hbar$/$\tau_0 \geq 2t_{a,c}$, the interchain hopping integrals
are {\it not} renormalized to zero as in the case of a strictly 1D system. Disordered Pr124 appears to be in a critical
phase of quasi-1D \lq frustrated metallicity' in which quasiparticle states remain extended over tens of unit cells and
local coherence co-exists with the development of localization corrections, albeit of an anomalous form
\cite{Narduzzo}, along the length of the crystal. One possibility is that this duality is macroscopic with large-scale
domains of alternating metallic and localized character. This would be surprising though given the stoichiometric
nature of Pr124. Moreover, the same localization threshold has been observed recently via electron irradiation
\cite{Rad06} which is known to produce only point-like defects. We therefore believe the duality is microscopic. One
might think of the chains as broken up into metallic islands separated by widely-spaced weak links between which the
quasiparticles remain long-lived and subject to an orbital Lorentz force. This picture bears certain similarities to
the so-called \lq interrupted strand' model (ISM) \cite{KuseZeller,RiceBernasconi} which was originally developed to
explain the observation of a Drude tail in the optical response of an apparently insulating organometallic complex
\cite{KuseZeller}.  In the ISM however, defects are assumed to be perfectly insulating (in the strictly 1D sense) with
no coherent tunneling between the interrupted segments. In Pr124, (occasional) coherent tunnelling between chains must
occur for the Lorentz force to be active. The key point of our proposed model is that in the regime where $\hbar/\tau_0
\sim t_{a,c}$ interchain hopping still remains a relevant perturbation, and electrons can still hop coherently, despite
not being the extended Bloch wavefunctions of the more metallic crystal. Crucially it seems disorder is more effective
at modifying the slope of $\rho_b(T)$ than it is at disrupting the electron's orbital motion. In this respect, the
behavior resembles more that of the \lq hopscotch' model of Phillips \cite{Phillips76} in which carriers can tunnel to
an adjacent chain via interchain scattering.

In conclusion we have uncovered an anomalously enhanced anisotropic T-MR in Pr124 crystals exhibiting localized
behavior that suggests charge carriers maintain an itinerant character over microscopic length scales. This unique
phenomenon occurs in Pr124 due to the low levels of disorder required to induce a dimensional crossover in the
electronic system and may therefore be absent in other quasi-1D metals where hopping integrals tend to be larger.
Whether aspects of TLL physics are manifest in Pr124 is still an interesting open question \cite{Take00, Mizo02},
though the strong violation of Kohler's rule in sample $b$2 is one possible indicator of such an exotic state.

We thank E. Arrigoni, J. R. Cooper and N. Shannon for helpful discussions and P. J. Heard, S. L. Kearns and M. M. J.
French for technical assistance. This work was supported by the EPSRC (U.K.)



\end{document}